# Understanding the Heart of the 5G Air Interface: An Overview of Physical Downlink Control Channel for 5G New Radio (NR)

Kazuki Takeda[‡], Huilin Xu[‡], Taehyoung Kim[§], Karol Schober[⊥], and Xingqin Lin[#]

[‡]Qualcomm Inc., [§]Samsung Electronics, [⊥]Nokia, [#]Ericsson Inc.

Emails: ktakeda@qti.qualcomm.com; huilinxu@qti.qualcomm.com; th86819.kim@samsung.com;
karol.schober@nokia-bell-labs.com; xingqin.lin@ericsson.com.

*Abstract*— New Radio (NR) is a new radio air interface developed by the 3rd Generation Partnership Project (3GPP) for the fifth generation (5G) mobile communications system. With great flexibility, scalability, and efficiency, 5G is expected to address a wide range of use-cases including enhanced mobile broadband (eMBB), ultra-reliable low-latency communications (URLLC), and massive machine type communications (mMTC). The physical downlink control channel (PDCCH) in NR carries Downlink Control Information (DCI). Understanding how PDCCH operates is key to developing a good understanding of how information is communicated over NR. This paper provides an overview of the 5G NR PDCCH by describing its physical layer structure, monitoring mechanisms, beamforming operation, and the carried information. We also share various design rationales that influence NR standardization.

## I. INTRODUCTION

THE 3rd Generation Partnership Project (3GPP) has developed the initial release of the fifth generation (5G) mobile communications system as 3GPP Release 15 [1]. New Radio (NR) is a radio access technology newly introduced in 3GPP Release 15. Though NR is not required to support backward compatibility with long-term evolution (LTE), ensuring harmonious LTE-NR coexistence has also been an important design consideration.

5G is expected to fulfill various requirements from a wide range of industries/verticals so that many use-cases such as enhanced mobile broadband (eMBB), ultra-reliable low-latency communications (URLLC), and massive machine type communications (mMTC), are supported [2-3]. In addition, 5G must support an extremely wide frequency range from below 1 GHz to above 50 GHz, which includes the frequencies that have not been utilized for cellular networks to date, e.g., millimeter waves [3]. New scenarios, use cases, requirements, and frequencies have posed many new challenges.

3GPP Release 15 NR is designed to address all these challenges. Some of the NR characteristics are highlighted below and further details can be found in [4].

(**1**) **Orthogonal frequency domain multiple access (OFDMA) with flexible subcarrier spacing (SCS)**: NR adopts OFDMA as its multiple access scheme. SCS values of $15 \times 2^\mu$ kHz with $\mu = 0, 1, 2, 3,$ or $4$ are supported [5]. The system selects SCS values based on carrier frequency, use-case, scenario, and/or requirements, and the SCS is configured using higher-layer signaling. Once the SCS is configured, frame structure is identified. Twelve subcarriers for a given SCS are grouped to form a physical resource block (PRB), which is used as a unit in frequency-domain resource allocation for channels/signals. Fourteen OFDM symbols for the given SCS are grouped to form a slot (a time unit commonly recognized among sets of user equipment (UEs) for the given SCS), where the slot duration is $2^{-\mu}$ ms. One or more slots comprise a 1 ms subframe, and 10 subframes are grouped to form a radio frame [5]. For carrier aggregation (CA), the SCS can be different across serving cells, i.e., carriers.

(**2**) **Flexible bandwidth**: NR is designed such that individual UEs can use smaller bandwidths than the system bandwidth. The UE bandwidth within a carrier, configured by the base station (BS) as the number of contiguous PRBs with an associated SCS is called a bandwidth part (BWP) [6]. Once the BWP is activated, data and control channels are received/transmitted within the BWP [7]. A UE can be configured with up to four BWPs, where BWPs may have different SCSs and may be mutually overlapping or non-overlapping in frequency. If more than one BWP is configured for a UE, the BS may select which BWP is active at a given time by means of downlink (DL) control. Therefore, the BS may adjust dynamically the UE bandwidth according to the amount/profile of data traffic for the UE. A reduction in the UE bandwidth may reduce the UE power consumption.

From the physical layer perspective, the data and signaling messages in NR are carried in the DL and uplink (UL) physical channels. Among these channels, the physical downlink control channel (PDCCH) plays a central role in, for example, DL scheduling assignments, UL scheduling grants, and special purposes such as slot format indication, preemption indication, and power control. Understanding how the PDCCH operates is key to developing a good understanding of how information is communicated over NR.

NR PDCCH has been studied and discussed in [5][9][10]. Performance study of wideband and narrowband precoding of NR PDCCH and a design of PDCCH hashing function can be found in [5]. The complexity analysis of NR PDCCH is presented in [9]. Brief overview of NR PDCCH is presented in [10]. In this article, we present a comprehensive A to Z overview of the final NR PDCCH design as well as we explain reasons that led to such design. This article presents the physical layer design of the NR PDCCH in Section II. Section III describes how a UE receives the NR PDCCH. In Section IV, beamforming operation of the NR PDCCH is explained. Section V provides details of the information carried by NR PDCCH. Section VI summarizes the article.

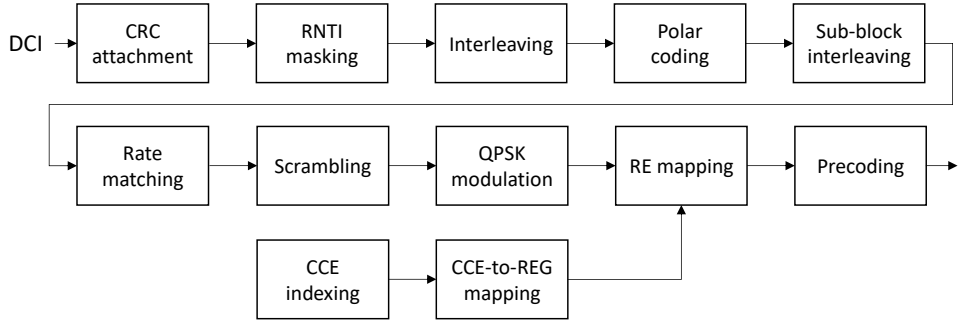

Fig. 1    Procedure for generating PDCCH from DCI.

## II. PDCCH PHYSICAL LAYER STRUCTURE

### A. Channel Coding and Downlink Control Information (DCI) Construction

The information carried by the PDCCH is referred to as DCI. DCI contains the scheduling information for the UL or DL data channels and other control information for one UE or a group of UEs. Further details regarding DCI are provided in Section V. DCI undergoes a sequence of processing steps to constitute the PDCCH payload.

The procedure for generating a PDCCH is illustrated in Fig. 1. If the size of the DCI format is less than 12 bits, a few zero padding bits will be appended until the payload size equals 12 bits [11]. For the DCI payload bits, a 24-bit cyclic redundancy check (CRC) is calculated and appended to the payload. The CRC allows the UE to detect the presence of errors in the decoded DCI payload bits. After the CRC is attached, the last 16 CRC bits are masked with the corresponding identifier, referred to as a radio network temporary identifier (RNTI). Using the RNTI mask, the UE can detect the DCI for its unicast data and distinguish sets of DCI with different purposes that have the same payload size. The CRC attached bits are then interleaved to distribute the CRC bits among the information bits. The interleaver supports a maximum input size of 164 bits. This means that DCI without CRC can have at most 140 of payload bits. The bits are then encoded by the Polar encoder to protect the DCI against errors during transmission. The encoder output is processed using a sub-block interleaver and then rate matched to fit the allocated payload resource elements (REs) of the DCI.

The payload bits of each DCI are separately scrambled by a scrambling sequence generated from the length-31 Gold sequence. The scrambling sequence is initialized by the physical layer cell identity of the cell or by a UE specific scrambling identity and a UE specific cell RNTI (C-RNTI). After the scrambled DCI bit sequence is Quadrature Phase Shift Keying (QPSK) modulated, the complex-valued modulation symbols are mapped to physical resources in units referred to as control channel elements (CCEs). Each CCE consists of six resource element groups (REGs), where a REG is defined as one PRB in one OFDM symbol which contains nine REs for the PDCCH payload and three demodulation reference signal (DMRS) REs. For each DCI, 1, 2, 4, 8, or 16 CCEs can be allocated, where the number of CCEs for a DCI is denoted as aggregation level (AL). With QPSK modulation, a CCE contains 54 payload REs and therefore can carry 108 bits. This requires the output size of the rate matching block to be $L \cdot 108$, where $L$ is the associated AL. Based on the channel environment and available resources, the BS can adaptively choose a proper AL for a DCI to adjust the code rate.

### B. Control Resource Sets

A DCI with AL $L$ is mapped to physical resources in a given BWP, where necessary parameters such as frequency and time domain resources, and scrambling sequence identity for the DMRS for the PDCCH are configured to a UE by means of control resource set (CORESET). In this subsection, we briefly introduce the CORESET, and then in Subsection II-C, we overview how the DCI with AL $L$ is mapped in the CORESET as a PDCCH.

A UE may be configured with up to three CORESETs on each of up to four BWPs on a serving cell. Therefore, a UE may be configured with up to 12 CORESETs on a serving cell in total, where each CORESET has an index of 0-11. In general, CORESETs are configured in units of six PRBs on a six PRB frequency grid (starting from a reference point referred to as point A) and one, two, or three consecutive OFDM symbols in the time domain. There is a special CORESET with index 0 (CORESET 0), which is configured using a four-bit information element in the master information block (MIB) with respect to the cell-defining synchronization signal and physical broadcast channel (PBCH) block (SSB). CORESET 0 can be acquired even before higher-layer configurations are provided, i.e., before additional system information or dedicated configuration is provided, while its configuration is restricted to a limited number of combinations of parameters compared to CORESETs with indices other than 0. Moreover, CORESET 0 may not be aligned with a six PRB frequency grid for other CORESETs, since the frequency-domain resource of CORESET 0 is determined relatively with the SSB as illustrated in Fig. 2 as an example, in which CORESET 0 is illustrated in red and CORESET 1 is illustrated in yellow on a BWP of 48 PRBs.

CORESETs are active only when their associated BWP is active, with the exception of CORESET 0, which is associated with the initial BWP (BWP with index 0). However, CORESET 0 can be monitored in another BWP when its numerology, i.e., SCS $\mu$ and cyclic prefix, matches the numerology of the non-initial BWP, and the PRBs of CORESET 0 are confined within the PRBs of the non-initial BWP.

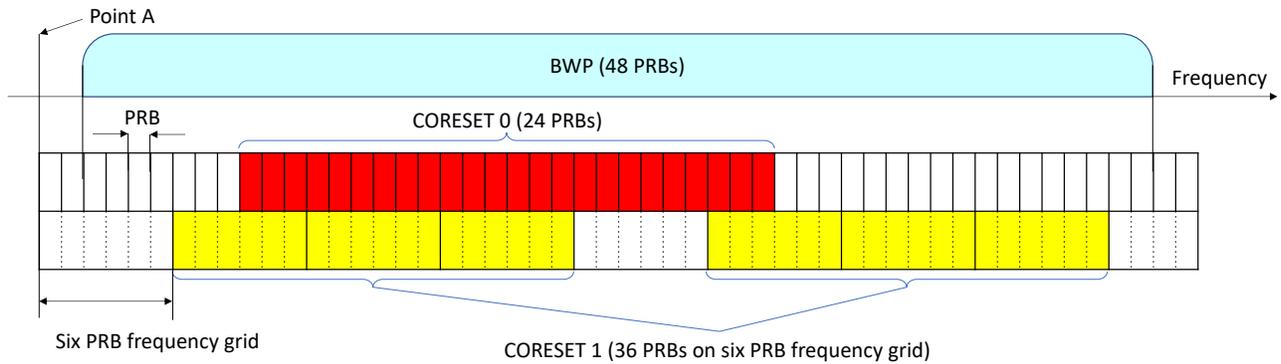

Fig. 2  Example of CORESET configurations on BWP.

*C. Mapping of PDCCH to CORESET*

A DCI of AL $L$ comprises $L$ continuously numbered CCEs, and the CCEs are mapped on a number of REGs in a CORESET.

NR supports distributed and localized resource allocation for a DCI in a CORESET. This is done by configuring interleaved or non-interleaved CCE-to-REG mapping for each CORESET. For interleaved CCE-to-REG mapping, REG bundles constituting the CCEs for a PDCCH are distributed in the frequency domain in units of REG bundles as illustrated in Fig. 3. A REG bundle is a set of indivisible resources consisting of neighboring REGs. A REG bundle spans across all OFDM symbols for the given CORESET. Therefore, interleaved CCE-to-REG mapping can enable both a time domain processing gain and frequency domain diversity. Interleaved CCE-to-REG mapping can be visualized as a process for which REG bundle indices are continuously filled in an array row-wise first and then read out column-wise. This process is often called block interleaving. By this means, adjacent CCEs for a PDCCH are broken down into scattered REG bundles in the frequency domain. On the other hand, for non-interleaved CCE-to-REG mapping, all CCEs for a DCI with AL $L$ are mapped in consecutive REG bundles of the CORESET.

Once the REGs corresponding to a PDCCH are determined, the modulated symbols of the PDCCH are mapped to the REs of the determined REGs in the frequency domain first and the time domain second, i.e. in increasing order of the RE index and symbol index, respectively.

NR supports both wideband and narrowband precoding for the PDCCH. In wideband precoding, PDCCH DMRSs are transmitted in all contiguous REGs of a CORESET carrying the PDCCH using the same precoder. On the other hand, in the narrowband precoding, PDCCH DMRSs are transmitted only in the REG bundles actually used for the PDCCH transmission, and precoding is constant only within the REG bundle. The former can maximize the frequency domain processing gain and the latter can take advantage of frequency dependent beamforming gain. Since the same precoding is applied to payload and the corresponding DMRS, precoding is transparent to the UE.

### III.  PDCCH MONITORING

In this section, we discuss how the PDCCH monitoring is configured to a UE and how the UE deals with the cases when the UE is configured with more CCEs or blind decodes than the UE is capable of monitoring.

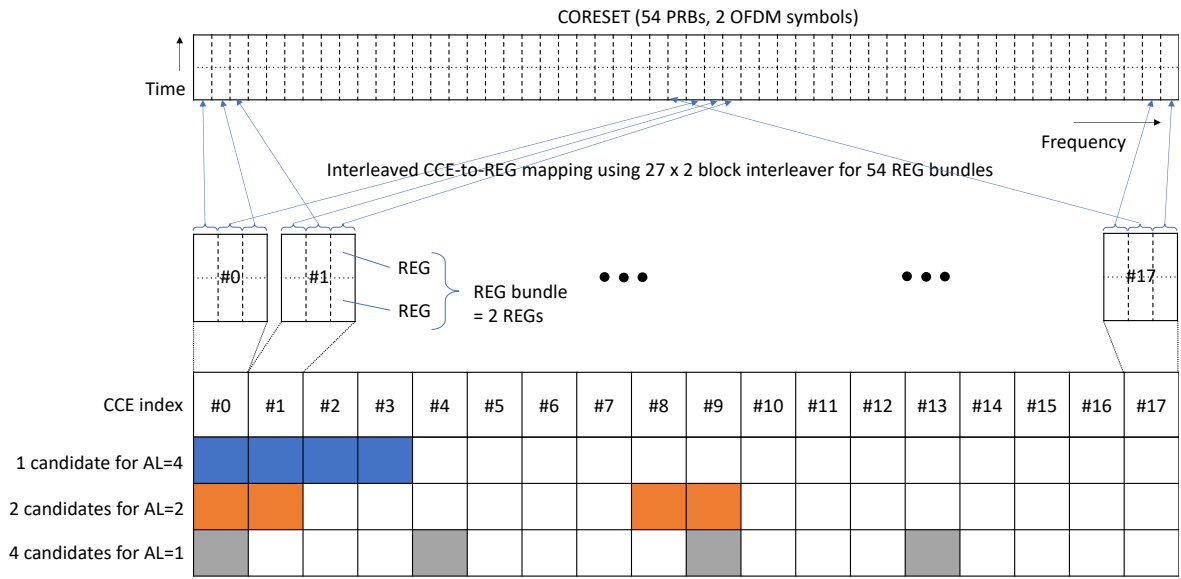

Fig. 3  Example of PDCCH candidate determination and REG mapping.

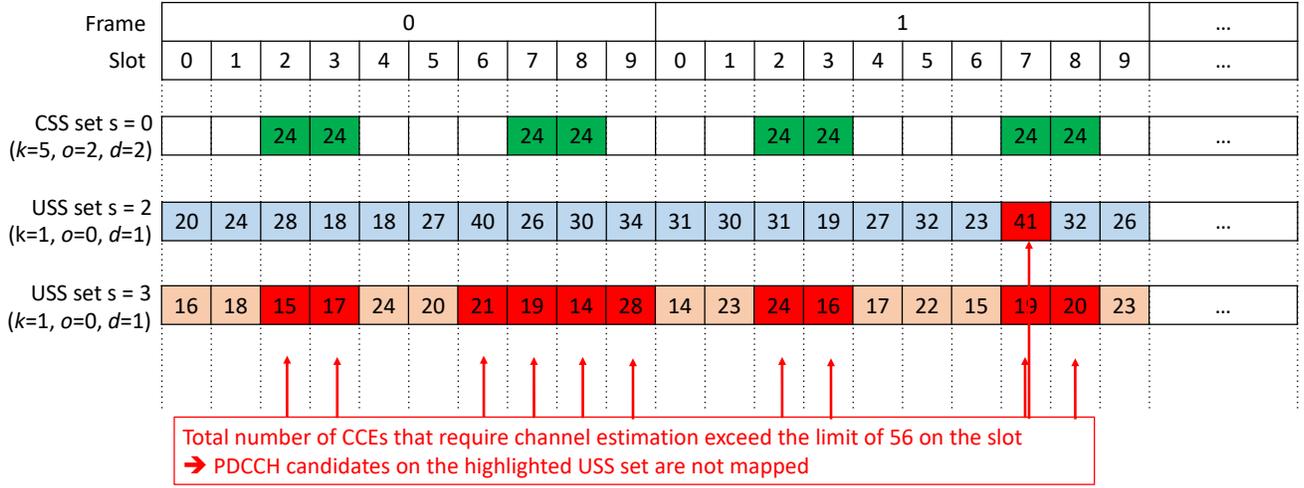

Fig. 4   Number of CCEs that require channel estimations.

*A.  Search Space Sets*

The UE performs blind decoding for a set of PDCCH candidates. PDCCH candidates to be monitored are configured for a UE by means of search space (SS) sets. There are two SS set types: common SS (CSS) set, which is commonly monitored by a group of UEs in the cell, and UE-specific SS (USS) set, which is monitored by an individual UE. A UE can be configured with up to 10 SS sets each for up to four BWPs in a serving cell. Therefore, a UE can be configured with up to 40 SS sets, where each has an index of 0-39. These numbers were selected sufficiently large to cover all scenarios envisioned by the industries so far. In general, SS set configuration provides a UE with the SS set type (CSS set or USS set), DCI format(s) to be monitored, monitoring occasion, and the number of PDCCH candidates for each AL in the SS set. There is a special CSS set with index 0 (SS set 0), which is configured using a four-bit information element in the MIB with respect to cell-defining SSB. The SS set 0 can be monitored even before higher-layer configurations are provided, while its configuration is restricted to a limited number of combinations of parameters compared to SS sets of other than 0.

A SS set with index *s* is associated with only one CORESET with index *p*. This mapping from a SS set to a CORESET enables simpler realization of multiple CORESETs and multiple SS sets in a cell. The UE determines the slot for monitoring the SS set with index *s* based on the higher layer parameters for periodicity *k*, offset *o*, and duration *d*, where periodicity *k* and offset *o* provide a starting slot and duration *d* provides the number of consecutive slots where the SS set is monitored starting from the slot identified by *k* and *o*.

Monitoring occasions of a SS set with index *s* within the slot are configured by a bitmap parameter in the SS set configuration. Each bit representing an OFDM symbol within the slot corresponds to the first OFDM symbol of the monitoring occasion of the SS set.

*B.  PDCCH Candidate Hash Function*

The mapping of PDCCH candidates of an SS set to CCEs of the associated CORESET is implemented by means of a hash function. The hash function randomizes the allocation of the PDCCH candidates within CORESET $p$ in slot $n_s$. This is done according to

$$L\left\{\left(Y_{p,n_s} + j_{p,m}^{(L)}\right) \bmod \lfloor N_{\text{CCE},p}/L \rfloor\right\} + i,$$

where a single carrier operation with a single SS set with index *s* is assumed for simplicity. The following are also assumed.
- *L* is the aggregation level,
- $N_{\text{CCE},p}$ is the total number of CCEs for given CORESET *p*,
- *m* (0, 1, …, $M^{(L)} - 1$) is the candidate index with $M^{(L)}$ being the number of PDCCH candidates for AL *L*,
- *i* (0, 1, …, $L - 1$) is the contiguous CCE index of the PDCCH candidate,
- $j_{p,m}^{(L)} = \lfloor \frac{m \cdot N_{CCE,p}}{L \cdot M^{(L)}} \rfloor$,
- $Y_{p,n_s} = 0$ for the CSS set,
- $Y_{p,n_s}$ (0, 1, …, $2^{16}$ - 1) for the USS set is a pseudo-random variable based on the C-RNTI of the UE and slot number $n_s$, and
- $\lfloor . \rfloor$ denotes the floor operation.

One example of an SS set determination in a CORESET, taking into account the CCE-to-REG mapping and the hash function, is illustrated in Fig. 3. The CORESET has 54 PRBs and two OFDM symbols with REG bundle size of two REGs. The SS set is assumed to have $M^{(L)} = (4, 2, 1, 0, 0)$ for AL $L =$ (1, 2, 4, 8, 16), respectively. For CCE-to-REG mapping, block interleaver with the row size of two is used. The NR hash function stems from LTE hash function design with higher configurability; other designs targeting further optimization, such as nested hash function from [5] or pseudo-random hash function from [9], was not adopted in NR.

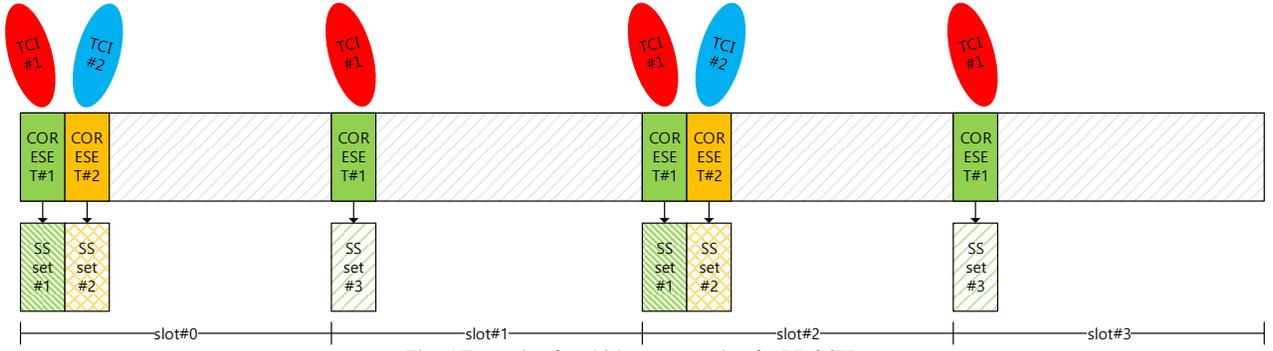
Fig. 5 Example of multi-beam operation for PDCCH

*C. UE Monitoring Capabilities and Overbooking*

Monitoring a large number of PDCCH candidates or CCEs increases the UE complexity. Therefore, NR specifies the maximum number of PDCCH candidates that require blind decodes, $M_{PDCCH}^{max,\mu} = \{44, 36, 22, 20\}$, and the maximum number of CCEs that require channel estimations, $C_{PDCCH}^{max,\mu} = \{56, 56, 48, 32\}$, that UEs are capable of handling in a slot of SCS $\mu = \{0, 1, 2, 3\}$. This limits the UE complexity to a reasonable level with an acceptable restriction on the configuration of SS sets for PDCCH monitoring.

When a UE is configured with more than one SS set, the number of PDCCH candidates/CCEs varies across slots due to independent monitoring occasions for the SS sets and slot-dependent hashing of different ALs for each SS set. Therefore, a base station (BS) is allowed to configure the UE with a number of PDCCH candidates/CCEs per slot that exceeds the UE capability, which is referred to as *overbooking*. Based on the configuration, the UE and BS map PDCCH candidates in each slot according to the following mapping rules: (i) CSS sets are mapped before USS sets, (ii) USS sets are mapped in ascending order of the SS set indices, and if the number of PDCCH candidates/CCEs exceeds either of the UE processing limits, then (iii) No more SS sets are mapped in the slot after reaching the UE processing limit. These rules allow BS to number UE's SS sets according to desired priority.

An example of mapping SS sets is illustrated in Fig. 4 assuming SCS $\mu = 0$. In this example, only the number of CCEs is illustrated for simplicity. We assume one CSS set $s = 0$, is configured consuming 24 constant CCEs in CORESET $p = 0$. CSS hashing does not depend on slot number $n_s$. In CORESET $p = 1$, two USS sets, $s = 2$ and $s = 3$, are configured, and their CCE consumption levels vary due to slot-dependent hashing of different ALs. When the total number of CCEs consumed by the CSS set and USS sets exceeds 56, the USS set $s = 3$ is not mapped. Without the overbooking feature, the configuration for SS sets would be restricted by the worst-case number and hence a BS is unable to use the UE monitoring capabilities to their maximum in all the slots. This results in potential multi-user candidate blocking.

*D. UE Monitoring Capabilities and Overbooking for CA*

Capabilities $M_{PDCCH}^{max,\mu}$ and $C_{PDCCH}^{max,\mu}$ defined in the previous subsection are referred to as a non-CA limit. In this subsection we introduce capabilities for CA, referred to as CA limit.

A BS configures a UE with $N_{cells}^{DL,\mu}$ serving cells, which does not exceed the UE capability of CA configuration. The UE monitors the PDCCH for each serving cell, where the non-CA limit for each serving cell still holds. However, for a CA with a large number of serving cells, the UE implementation complexity or power consumption for PDCCH monitoring may be problematic. Therefore, NR allows the UE to report the limits for CA (CA-limit), which is less than the non-CA limit multiplied by the number of configured serving cells. In particular, a UE may report the number of serving cells that it is capable of monitoring $N_{cells}^{cap} \geq 4$. This means that when a BS configures the number of serving cells $N_{cells}^{DL,\mu} \leq 4$ or the UE is capable of monitoring more cells than the BS configures $N_{cells}^{DL,\mu} \leq N_{cells}^{cap}$, the UE capability is a non-CA limit on each of the serving cells. For the remaining case when $4 < N_{cells}^{cap} < N_{cells}^{DL,\mu}$, the CA-limit is applied. The BS takes into account the UE capability to configure SS sets for CA.

Note that the UE reports a single value for $N_{cells}^{cap}$, even if the UE supports CA with various SCS combinations across serving cells where the UE monitors PDCCH candidates. For CA with different SCSs across serving cells, the serving cells are grouped based on their SCS $\mu$, and the total number of PDCCH candidates that require blind decoding is split into the groups of cells of the same SCS. In particular, the maximum number of PDCCH candidates that requires blind decoding, per group of cells of SCS $\mu$, i.e. the CA-limit, is defined as

$$M_{PDCCH}^{total,\mu} = \lfloor N_{cells}^{cap} M_{PDCCH}^{max,\mu} N_{cells}^{DL,\mu} / \sum_{j=0}^{3} N_{cells}^{DL,j} \rfloor.$$

Similarly, NR defines the CA-limit for the maximum number of CCEs that require channel estimations per group of cells of SCS $\mu$, as

$$C_{PDCCH}^{total,\mu} = \lfloor N_{cells}^{cap} C_{PDCCH}^{max,\mu} N_{cells}^{DL,\mu} / \sum_{j=0}^{3} N_{cells}^{DL,j} \rfloor.$$

Within a group of cells of the same numerology $\mu$, the limits, $M_{PDCCH}^{total,\mu}$ and $C_{PDCCH}^{total,\mu}$, can be shared among cells, however, still subject to non-CA limit per cell. Finally, the overbooking limit is defined as the minimum between CA-limit and non-CA-limit as $min(M_{PDCCH}^{total,\mu}, M_{PDCCH}^{max,\mu})$ and $min(C_{PDCCH}^{total,\mu}, C_{PDCCH}^{max,\mu})$. Note that for CA, overbooking is allowed only on the primary cell of the cells group where in general various SS sets, e.g., CSS sets and USS sets, are monitored. For a secondary cell where relatively fewer SS sets are monitored, the BS ensures that the maximum numbers of PDCCH candidates and non-overlapping CCEs do not exceed the overbooking limits.

## IV. BEAMFORMING OPERATION

To overcome serious propagation loss and blockage for high frequency ranges such as millimeter wave, NR can utilize highly directional beamformed transmission and reception using a large number of antennas between the BS and UE [12]. Hybrid beamforming is a combination of analog beamforming that applies different phase shifters and/or amplitude weights on each antenna panel and digital beamforming that applies different digital precoders across panels. Since the analog beamforming can only cover a limited area due to a narrow beam width, multiple beams need to be utilized to cover the entire cell.

To support multi-beam based operation for the PDCCH, NR supports a higher-layer configuration for beamforming (transmission configuration indication (TCI) state configuration) per CORESET. When a UE monitors a SS set associated with a CORESET, the UE can receive the PDCCH in the CORESET based on the TCI state configuration configured for the CORESET as illustrated in Fig. 5.

In the following subsections, we discuss more details on the PDCCH beamforming operation.

### A. Beam Configuration Aspects

Beam information for PDCCH reception is implicitly known to the UE by means of the quasi-co-location (QCL) relationship between DL reference signals (RSs) (typically channel state information RS (CSI-RS), each of which is associated with each beam) and the DMRS of the PDCCH. In NR, the DMRS of the PDCCH can be quasi co-located with a RS with QCL-TypeA and/or QCL-TypeD. QCL-TypeA corresponds to the channel statistical properties observed on the UE side including Doppler shift, Doppler spread, average delay, and delay spread. QCL-TypeD corresponds to the receiver beam information on the UE side. If the DMRS of the PDCCH is quasi co-located with a RS with QCL-TypeD, this means that the UE may use the same spatial reception parameters used for receiving the RS in analog beamforming to receive the PDCCH.

The BS can explicitly configure the QCL relationships using higher layer signaling. The UE can be configured with up to 64 TCI states for a CORESET to receive a PDCCH. Each TCI state contains parameters on RS resources and the QCL relationship between the RS(s) and the DMRS ports of the PDCCH with respect to QCL-TypeA and QCL-TypeD. To receive PDCCH at a time, only one beam is used. Therefore, if more than one TCI state is configured for a CORESET, the BS activates one of the TCI states used for the CORESET using a medium access control (MAC) control element (CE) activation command [13].

If a TCI state has not been indicated to a UE, the UE monitors the PDCCH in monitoring occasions of SS sets using the receiver beam that the UE used for receiving the SSB selected during the initial access procedure. In other words, the UE can assume that the DMRS antenna port associated with the PDCCH in the CORESET is quasi co-located with the corresponding SSB with respect to QCL-TypeA and QCL-TypeD properties.

### B. Beam Failure Recovery Aspects

NR supports link recovery procedure based on layer 1 reference signal received power (L1-RSRP) measurement, which is referred to as the beam failure recovery (BFR) procedure. That is, once there is a misunderstanding regarding the appropriate beams between the BS and UE that leads to poor link quality, the UE can indicate a preferred beam to the BS by transmitting a physical random access channel (PRACH) on the resource associated with the preferred beam.

Table 1 Summary of DCI formats in 5G NR

| DCI format | Usage | RNTI for CRC scrambling |
|---|---|---|
| DCI 0_0 | Fallback DCI format for scheduling of PUSCH | • C-RNTI or MCS-C-RNTI for dynamically scheduled unicast transmission<br>• CS-RNTI for configured scheduled unicast transmission<br>• TC-RNTI for Message 3 transmission in contention based random access |
| DCI 0_1 | Non-fallback DCI format for scheduling of PUSCH | • C-RNTI or MCS-C-RNTI for dynamically scheduled unicast transmission<br>• CS-RNTI for configured scheduled unicast transmission<br>• SP-CSI-RNTI for activation of semi-persistent channel state information reporting on PUSCH |
| DCI 1_0 | Fallback DCI format for scheduling of PDSCH | • C-RNTI or MCS-C-RNTI for dynamically scheduled unicast transmission<br>• CS-RNTI for configured scheduled unicast transmission<br>• SI-RNTI for broadcasting of system information<br>• P-RNTI for paging and system information change notification<br>• RA-RNTI for random access response<br>• TC-RNTI for contention resolution (when no valid C-RNTI is available) |
| DCI 1_1 | Non-fallback DCI format for scheduling of PDSCH | • C-RNTI or MCS-C-RNTI for dynamically scheduled unicast transmission<br>• CS-RNTI for configured scheduled unicast transmission |
| DCI 2_0 | Notifying a group of UEs of the slot format | • SFI-RNTI for slot format indication |
| DCI 2_1 | Notifying a group of UEs of the PRB(s) and OFDM symbol(s) where UE may assume no transmission is intended for the UE | • INT-RNTI for pre-emption indication in downlink |
| DCI 2_2 | Transmission of a group of power control commands for PUCCH and PUSCH | • TPC-PUCCH-RNTI for PUCCH power control<br>• TPC-PUSCH-RNTI for PUSCH power control |
| DCI 2_3 | Transmission of a group of SRS requests and power control commands for SRS transmissions | • TPC-SRS-RNTI for SRS trigger and power control |

A UE can be configured with a set of CSI-RS resources (a set $q_0$) for the purpose of beam failure detection. Then the UE measures the link quality according to the CSI-RS resources in set $q_0$. If the measured radio link quality for all corresponding CSI-RS resources in set $q_0$ is worse than the pre-defined threshold, the UE can request BFR by transmitting the PRACH corresponding to the selected $q_{new}$ from the candidate SSB or CSI-RS set $q_1$ which is configured for BFR. After transmitting the PRACH, the UE monitors PDCCH in a search space configured for BFR (SS-BFR) to receive the corresponding response. Once a PDCCH with the response is successfully received from the SS-BFR, the UE considers that the BFR is successful.

During the PDCCH monitoring in the SS-BFR, the UE assumes that the PDCCH DMRS is quasi co-located with the selected CSI-RS or SS/PBCH block with index $q_{new}$. The UE can receive the physical downlink shared channel (PDSCH) (scheduled by the PDCCH transmitted through the SS-BFR) providing the MAC CE activation command for TCI state update/change for the CORESETs.

*C. Beam Collision Aspects*

The number of analog beams that can be received simultaneously is restricted by the number of antenna panels supported on the UE side since analog beamforming is applied per antenna panel. In NR Release 15, the study focused on a single panel scenario, so that a UE can receive the physical channels at the same time using a single analog beam. Taking into account this restriction, some PDCCH monitoring rules are defined in NR.

When some PDCCH monitoring occasions in multiple CORESETs configured with different QCL-TypeD overlap in the time domain, the UE monitors PDCCHs only in the CORESET configured to monitor the CSS set with the lowest index. If all the overlapped PDCCH monitoring occasions are not associated with any CSS set, the PDCCHs in the CORESET configured with the USS set with the lowest index are monitored.

## V. DOWNLINK CONTROL INFORMATION

In this section, we discuss the details of DCI which is the payload transmitted on a PDCCH.

A UE identifies the payload size and RNTI used for CRC scrambling for a DCI format as presented in Section II. There are various DCI formats and RNTIs, as summarized in Table 1.

Downlink scheduling assignments are supported by DCI format 1_0 (known as the fallback format) and DCI format 1_1 (known as the non-fallback format). The design rationale is that the fallback format as a default scheduling option should be robust to support the basic NR operations while the non-fallback format should be flexible to accommodate the rich set of NR features. Specifically, DCI format 1_0 supports a basic set of NR functionality and its fields are generally not configurable. DCI format 1_1 supports all NR features and its fields are highly configurable. In general, DCI format 1_0 is smaller than DCI format 1_1. DCI format 1_0 is particularly needed for scheduling system information, paging and system information change notification, random access response, and contention resolution (when no valid C-RNTI is available).

Uplink scheduling grants are supported by DCI format 0_0 (known as the fallback format) and DCI format 0_1 (known as the non-fallback format). Similar to the DCI formats used for DL scheduling assignments, DCI format 0_0 supports a basic set of NR functionality while DCI format 1_1 supports all NR features.

In addition to the DCI formats used for DL scheduling assignments and UL scheduling grants, there are DCI formats 2_0, 2_1, 2_2, and 2_3 introduced for special purposes of signaling to a configured group of one or more devices. The CRCs of these DCI formats are scrambled by specific RNTIs, as described in Table 1. DCI format 2_0 is used for slot format indication that dynamically determines whether the resources are used for UL or DL transmission. DCI format 2_1 is used for preemption indication that signals that some resources have been preempted and thus are not used for transmission to the notified device(s). DCI format 2_2 is used for transmitting physical uplink shared channel (PUSCH) or physical uplink control channel (PUCCH) power-control commands. The main use case is to support power control for semi-persistent or periodic uplink transmissions. DCI format 2_3 is used for transmitting sounding reference signal (SRS) power-control commands and optionally SRS requests.

Different DCI formats may or may not have the same DCI size. An NR UE is capable of monitoring up to three different DCI sizes using the C-RNTI. Additionally, the UE is capable of monitoring one additional DCI size using RNTIs for special purposes (such as SFI-RNTI and INT-RNTI). The DCI scrambled with the C-RNTI (MCS-C-RNTI and CS-RNTI) is time critical and requires the UE to decode it promptly to process the scheduled data transmission. The DCI scrambled with a RNTI used for a special purpose is less time-critical for the UE to decode. This requirement is usually referred to as a "3+1" DCI size budget.

Due to the constraint of the DCI size budget, the sizes of some DCI formats need to be aligned by padding, truncation, and/or determining the frequency domain resource assignment field differently.

## VI. PDCCH ENHANCEMENTS IN RELEASE 16

The first 5G NR specifications were completed in Release 15. 3GPP has started to evolve NR in Release 16 [14] to further improve performance and address new use cases. PDCCH enhancements are an important part of Release 16 and span across multiple topics, as overviewed below. In Release 16, physical layer enhancements for NR URLLC is being specified. For some URLLC services, frequent PDCCH monitoring and/or reliability enhancements to PDCCH are required to meet the latency and reliability requirements. 3GPP Release 16 is considering to enhance the PDCCH in terms of both latency and reliability viewpoints. As for the latency enhancements, PDCCH monitoring capability can be re-defined and upgraded; it can change from "per slot" to "per monitoring span" and specify higher values for Release 16 compared to Release 15. As for the reliability enhancements, various DCI fields for PDCCH would become variable fields. With this, DCI message payload can be controlled by network such that it can be very compact and hence, the system can take a good balance between higher performance and higher flexibility.

In vehicular-to-everything (V2X), the Physical Sidelink Control Channel (PSCCH) is defined as the counterpart of PDCCH for control information transmission between UEs. Correspondingly, the Sidelink Control Information (SCI) is defined as the counterpart of DCI for PDCCH. Unlike PDCCH, PSCCH supports a 2-stage control channel design. Decoding of the first stage SCI requires blind detection. The first stage SCI only needs to carry the basic information such as resource allocation for the data channel and the second stage SCI. This design seems less flexible than PDCCH but it provides enough flexibility for the broadcast operation of the sidelink. The second stage SCI carries additional information that is not used for scheduling and it does not need blind detection. With this design, the number of blind detections for the first stage PSCCH decoding can be less than the PDCCH for which separate blind detections are needed for DCI formats with different sizes. As a result, the processing complexity of PSCCH decoding can be reduced.

In NR unlicensed (NR-U) in sub-7GHz, a BS monitors the wireless channel and checks whether it has gained the right to use the channel based on Listen Before Talk (LBT) mechanism. LBT is performed in granularity of 20 MHz sub-bands. Once LBT passes for a sub-band, the BS can transmit (PDCCH, data, etc.) on the sub-band to UE(s). When UE's BWP comprises multiple 20 MHz sub-bands, specifications allow to confine a PDCCH within an LBT sub-band to avoid partial puncturing of PDCCH candidates when LBT fails for some sub-bands of the BWP. Confinement is enabled by a configuration of a NR-U CORESET within a sub-band that can be replicated in the sub-bands of the BWP, as if the same CORESET would be separately configured in those sub-bands. This reduces the complexity of handling additional CORESETs, as these CORESETs share the same set of parameters and the same QCL assumption. Additionally, the configuration overhead is reduced. NR-U will likely use the NR Release 15 DCI formats enhanced with NR-U specific fields to transmit control information. For example, an additional field is added to NR DCI format with CRC scrambled by SFI-RNTI to indicate sub-bands of a BWP on which the BS is transmitting.

Enhanced MIMO in Release 16 includes support of multiple transmit receive point (multi-TRP) transmission. In multi-TRP, different data from multiple TRPs may be transmitted for spatial multiplexing to increase data rate, or same data from multiple TRPs may be transmitted for diversity to improve transmission reliability and robustness. To efficiently support multi-TRP, both single PDCCH and multiple PDCCH based schemes are being introduced in Release 16. In single PDCCH based scheme, one TRP sends PDCCH to schedule one set of PDSCH layers from a first TRP and a second set of PDSCH layers from a second TRP. In multiple PDCCH based scheme, two TRPs independently schedule PDSCHs from two TRPs.

From UE power saving perspective, power saving techniques to reduce the UE's PDCCH monitoring operations is under discussion in Release 16. It was shown that a large amount of UE's power in modem composed of radio frequency (RF) chain and base band (BB) unit is consumed due to the PDCCH monitoring without any scheduled data [15]. Motivated by this, a power saving signal (a.k.a wake-up signal) is introduced to dynamically control the UE's PDCCH monitoring behavior depending on the data traffic. The power saving signal is transferred by PDCCH of which the monitoring occasions are located before discontinuous reception (DRX) cycle. The power saving signal can indicate whether the UE skips the subsequent monitoring occasions within the DRX on duration or not. By this way, the UE will wake up to monitor PDCCH only when there is scheduled data. In [15], it was evaluated that a remarkable power saving gain can be achieved with endurable latency loss under the low data traffic circumstance.

## VII. CONCLUSIONS

This article presented an overview of the 5G NR PDCCH with details on the mechanisms including physical layer structure, monitoring schemes, beamforming operation, and the carried information. The design of the 5G NR PDCCH enables various use-cases and operations in a wide range of carrier frequencies.

## REFERENCES

[1] http://www.3gpp.org/news-events/3gpp-news/1965-rel-15_news, Jan. 2019.
[2] 3GPP TS22.261, "Service requirements for 5G system; Stage 1," v15.0.0, Mar. 2017.
[3] 3GPP TR38.913, "Study on scenarios and requirements for next generation access technologies," v15.0.0, July 2018.
[4] X. Lin, *et al*., "5G new radio: Unveiling the essentials of the next generation wireless access technology," *arXiv preprint arXiv:1806.06898*, to appear in IEEE Communications Standards Magazine, June 2018.
[5] 3GPP TS38.211, "NR; Physical channels and modulation," v15.3.0, Sept. 2018.
[6] 3GPP TS38.213, "NR; Physical layer procedure for control," v15.3.0, Oct.2018.
[7] 3GPP TS38.214, "NR; Physical layer procedure for data," v15.3.0, Oct.2018.
[8] Fatemeh Hamidi-Sepehr; *et al.*, "5G NR PDCCH: Design and Performance", 2018 IEEE 5G World Forum (5GWF), July 2018
[9] Volker Braun, *et al.*, "5G NR Physical Downlink Control Channel: Design, Performance and Enhancements", IEEE Wireless Communications and Networking Conference, Apr. 2019
[10] Erik Dahlman et al. " 5G NR: The Next Generation Wireless Access Technology", Academic Press, Aug 2018
[11] 3GPP TS38.212, "NR; Channel coding and multiplexing," v15.3.0, Sept. 2018.
[12] E. Onggosanusi *et al*, "Modular and high-resolution channel state information and beam management for 5G new radio," *IEEE Communications Magazine*, vol. 56, no 3, pp. 48 -55, March 2018.
[13] 3GPP TS38.321, "NR; Medium Access Control (MAC) protocol specification" v15.3.0, Sept. 2018.
[14] 3GPP TR 21.916, "Summary of Rel-16 Work Items," v0.1.0, Sept. 2019.
[15] 3GPP TR 38.840, "Study on UE Power Saving," v2.0.0, May 2019.